\documentclass[aps,prl,superscriptaddress,amsfonts,amsmath,amssymb,showpacs,floatfix,reprint,longbibliography,footinbib]{revtex4-2}

\usepackage{url}
\usepackage{bm,bbm}
\usepackage{graphicx}
\usepackage{color}
\usepackage[colorlinks=true, urlcolor=blue, linkcolor=blue, citecolor=blue, pdftex]{hyperref}
\usepackage[english]{babel}
\usepackage{physics}

\newcommand{\T}{\widetilde{T}}

\usepackage{soul}
\usepackage{blindtext}
\usepackage{lipsum}
\usepackage{nicefrac}

\newcommand{\dagga}{{\phantom{\dagger}}}

\begin{document}

\title{Stability of algebraic spin liquids coupled to quantum phonons}

\author{Francesco Ferrari}
\thanks{These two authors contributed equally to this work}
\affiliation{Institute for Theoretical Physics, Goethe University Frankfurt, Max-von-Laue-Stra{\ss}e 1, D-60438 Frankfurt a.M., Germany}

\author{Josef Willsher}
\thanks{These two authors contributed equally to this work}
\affiliation{Technical University of Munich, TUM School of Natural Sciences, Physics Department, 85748 Garching, Germany}
\affiliation{Munich Center for Quantum Science and Technology (MCQST), Schellingstr. 4, 80799 München, Germany}

\author{Urban F. P. Seifert}
\affiliation{Kavli Institute for Theoretical Physics, University of California, Santa Barbara, CA 93109}
\affiliation{Institute for Theoretical Physics, University of Cologne, 50937 Cologne, Germany}

\author{Roser Valent\'\i}
\affiliation{Institute for Theoretical Physics, Goethe University Frankfurt, Max-von-Laue-Stra{\ss}e 1, D-60438 Frankfurt a.M., Germany}

\author{Johannes Knolle}
\affiliation{Technical University of Munich, TUM School of Natural Sciences, Physics Department, 85748 Garching, Germany}
\affiliation{Munich Center for Quantum Science and Technology (MCQST), Schellingstr. 4, 80799 München, Germany}
\affiliation{Blackett Laboratory, Imperial College London, London SW7 2AZ, United Kingdom}

\date{\today}

\begin{abstract}
Algebraic spin liquids are quantum disordered phases of insulating magnets which exhibit fractionalized gapless excitations and power-law correlations. 
Quantum spin liquids in this category include the experimentally established 1D Luttinger liquid, as well as the U(1) Dirac spin liquid (DSL) which has been a focus of recent candidate materials searches. Most notably, several exchange-frustrated Heisenberg materials on the triangular lattice have shown evidence of the U(1) DSL.
In this work, we measure the algebraic correlations of spin-singlet excitations in the $J_1$--$J_2$ antiferromagnetic Heisenberg model on the triangular lattice, prompting a detailed investigation of this model's stability under spin-phonon coupling using variational Monte Carlo. As seen before in 1D spin chains, we observe a low-temperature transition from a U(1) DSL to valence bond order and predict the parameter regime where the model realizes a stable DSL ground state.
To achieve this, we employ a series of finite-size scaling Ansätze inspired by the low-energy DSL's conformal description in terms of quantum electrodynamics, and show that emergent monopole operators drive the instability. We compare the physics of this transition to the 1D Luttinger liquid throughout our analysis. We derive the regime of stability against spin-Peierls ordering and argue that the DSL ground state might still be achievable in candidate materials, despite its tendency to valence bond solid ordering.
\end{abstract}

\maketitle

\paragraph{Introduction.}
Quantum spin liquids (QSLs) are exotic phases of matter characterized by fractionalized excitations and a long-ranged entanglement structure, possibly realised in magnetic systems where the spins avoid magnetic ordering down to zero temperature due to frustration effects~\cite{knolle2019,broholm2020}.
QSLs and their dynamics evade a description in terms of a Landau framework of order parameters for symmetry-broken states~\cite{wen2002}. Instead, they can be portrayed as deconfined phases of emergent gauge theories, which make the fractionalization of spins into \emph{spinon} quasiparticles, coupled to emergent dynamical gauge fields~\cite{savary2016}, manifest. Assessing the intrinsic stability of these states as well as their robustness against external perturbations is a challenging task, but remains a critical question on the road to their realization in materials.

The focus of this letter is the U(1) Dirac spin liquid (DSL) state, where this challenge is complicated further by the strongly interacting nature of the appropriate emergent gauge theory, quantum electrodynamics in 2+1 dimensions (QED$_3$) \cite{wen2002,hermele2004}: at low energies, this theory of gapless Dirac fermions (spinons) interacting with a dynamical compact U(1) gauge field is believed to flow to an interacting fixed point with conformal symmetry \cite{song2019}.
Recently, a wealth of numerical evidence has accumulated that the U(1) DSL could be realized in the frustrated $J_1$--$J_2$ Heisenberg model on the triangular lattice, as an effect of the competition of first- and second-neighbor antiferromagnetic exchange. 
Different numerical methods reported a stable, extended spin-liquid regime around the highly frustrated point $J_2/J_1 = 1/8$~\cite{kaneko2014,zhu2015,hu2015,iqbal2016,hu2019}.
Most importantly, spectral functions within the QSL phase show indication of Dirac spinon excitations~\cite{ferrari2019,sherman2023,drescher2023} and energy-level spectroscopy yields evidence for low-lying monopole states~\cite{wietek2023}, consistent with QED$_3$.

The emergent conformal symmetry implies that correlation functions must have a power-law decay ${C_i(\bm{r}) \propto \| \bm{r}\|^{-2\Delta_i}}$ 
at large distances.
The emergent SU(4) symmetry of the low-energy theory requires that this exponent $\Delta_\Phi$ must be the same for both spin-spin and dimer-dimer correlations (at appropriate wavevectors), and is given by the scaling dimension of the monopole $\Phi$.
This power-law behaviour
motivated the name `algebraic' QSL~\cite{rantner2001,hermele2004,hermele2005,hermele2008}, suggesting an analogy with the gapless spin-liquid ground state of the spin-half Heisenberg chain \cite{song2019,mcgreevy2019}, which is predicted to have power-law correlations controlled by conformal data and a similar emergent SO(4) symmetry \cite{PhysRevLett.95.036402}.
Such critical correlations of the 1D spin liquid can render it unstable to perturbations; dimer-dimer correlations couple to lattice distortions to precipitate a spin-Peierls instability towards a valence-bond solid (VBS) dimerized state, and spin-spin correlations lead to N\'eel ordering under weak interchain coupling, as may appear in a bulk 3D sample \cite{giamarchi2003}.
In 1D, the prediction of correlation functions in gapless spin liquids and the properties of the corresponding instabilities are facilitated both by powerful analytical frameworks (such as bosonization and the Bethe Ansatz), and accurate numerical tools, like the density matrix renormalization group (DMRG)~\cite{bursill1999}. 

While quasi-1D spin liquid behaviour is well established experimentally~\cite{mourigal2013,lake2013}, candidate materials in 2D are less well characterized. 
Recently however, several candidates have also been identified in 2D which show signs of a gapless ground state and are potentially predominantly described by the triangular lattice $J_1$--$J_2$ model, such as YbZn$_2$GaO$_5$~\cite{xu2023} and Ytterbium-delafossite compounds $A$YbSe$_2$, where $A$ is an Alkali metal~\cite{xie2023,scheie2024,scheie2024b}. 
A theoretical challenge is understanding the stability of the 2D U(1) DSL in more realistic models, which account for the experimental realities such as disorder or spin-lattice coupling.
This is particularly pressing, given the algebraic nature of the U(1) DSL and recent field-theory based predictions of a weak-coupling instability when coupled to (classical) phonon modes~\cite{seifert2024} or in bilayers~\cite{luo2022}.

\begin{figure}
\includegraphics[width=0.48\textwidth]{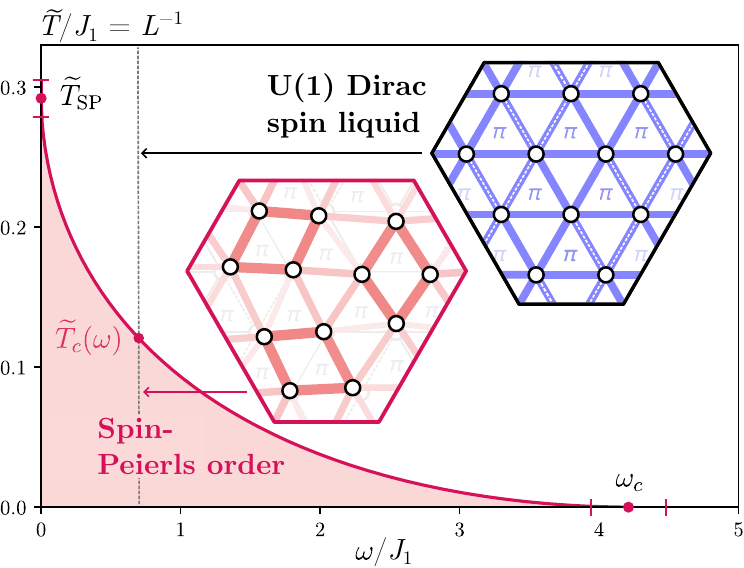}

\caption{\label{fig:phasediag} Phase diagram of the frustrated triangular lattice Heisenberg model coupled to dynamical phonons with energy $\omega$, for fixed  spin-lattice coupling $\lambda=0.3$. Below $\omega_c$, we measure a finite-temperature phase transition $\T_c(\omega)$ from the U(1) Dirac spin liquid to spin-Peierls order (i.e., simultaneous valence bond solid order and lattice distortion).
We use a variational method, where the spin wave function in the liquid regime is the fermionic state with alternating $0$ and $\pi$ fluxes (see upper-right inset for our gauge choice: on dotted bonds, the fermion hoppings have a negative sign)~\cite{lu2016}. The strengths of the $\langle S^z_i S^z_j\rangle$  correlations of nearest-neighbors are shown in the insets through the thickness/shading of the bonds.
Finite-temperature responses are inferred from the finite-size behaviour of the system. We find that a conformal scaling Ansatz well describes the scaling of critical temperature, using which we measure the scaling dimension of monopoles $\Delta_\Phi=1.23(2)$.
}
\end{figure}

In this Letter, we study the stability of the DSL in the triangular lattice $J_1$--$J_2$ model when spins are coupled to phonons.
We use a series of variational Monte Carlo (VMC) calculations to study the model's dimer-dimer correlations and argue to find good evidence for monopoles dominating the algebraic response.
We then assess the behaviour of the spin-Peierls transition, considering both the classical limit of static distortions, as well as a full dynamical treatment of quantum phonons. 
We find a U(1) DSL--VBS transition that is well described by CFT predictions: the critical spin-lattice coupling goes to zero in the limit of adiabatic phonons, but remains finite when the phonon energy scale $\omega$ is large.
A series of scaling collapses inspired by the conformal data of the QED$_3$ continuum theory show a good agreement with monopole scaling dimension $\Delta_\Phi\approx1.25$ predicted from analyses of pure QED$_3$ with $N_f=4$ fermions \cite{10.21468/SciPostPhys.13.2.014,PhysRevD.105.085008,PhysRevD.100.054514}.
The induced VBS order has a 12-site unit cell and the scaling of the measured critical spin-lattice coupling is described by a power-law $\lambda_c\sim\omega^{3-2\Delta_\Phi}$.
We plot our predictions of the coupled spin-lattice phase diagram in Fig.~\ref{fig:phasediag} for a moderate coupling strength $\lambda=0.3$ (on the order seen in 1D systems \cite{huizinga1979,bray1975,hase1993}). 
Together, this shows that a finite-temperature spin-Peierls phase transition away from a liquid state may be expected in 2D as a direct result of the emergent monopole operators in the low-energy theory.

\paragraph{Model and correlations.}
The starting point of our study is the $S=1/2$, $J_1$--$J_2$ Heisenberg model on the triangular lattice
\begin{align}
{\cal H}_{J_1 J_2} &=J_1 \sum_{\langle i,j\rangle} \vec{S}_i \cdot \vec{S}_{j} + J_2 \sum_{\langle \langle i,j\rangle \rangle} \vec{S}_i \cdot \vec{S}_{j} 
\label{eq:spin_hamiltonian}
\end{align}
with $J_1,J_2>0$. According to different numerical approaches, the phase diagram of the model features an extended QSL region for ${0.07(1)\lesssim
 J_2/J_1 \lesssim
0.16(1)}$~\cite{kaneko2014,zhu2015,hu2015,iqbal2016,hu2019}.
A variational description of the QSL can be formulated within the Abrikosov representation, in which spin operators are expressed in terms of fermionic partons, ${\vec{S}_i = \frac{1}{2} f_{i,\alpha}^\dagger\vec{\sigma}_{\alpha\beta}f^\dagga_{i,\beta}}$, subjected to the one-fermion-per-site constraint~\cite{wen2002}.
Following the projective symmetry group classification of QSL phases on the triangular lattice~\cite{lu2016}, the VMC study of Ref.~\cite{iqbal2016} showed that the lowest-energy variational Ansatz is a U(1) DSL, defined by the fermionic Hamiltonian
\begin{equation}\label{eq:H0fermions}
{\mathcal{H}}_{\rm 0} = \sum_{\langle i,j \rangle} \sum_{\alpha=\uparrow,\downarrow} t_{i,j} f^\dagger_{i,\alpha} f^\dagga_{j,\alpha} + {\rm h.c.} \ \ .
\end{equation}
The nearest-neighbor hoppings $t_{i,j} = t e^{i a_{i,j}}$ produce alternating $0$ and $\pi$ fluxes through adjacent triangular plaquettes (see Fig.~\ref{fig:phasediag}), and the U(1) gauge redundancy is given by $f^\dagger_{i,\alpha} \to e^{i\theta_i}f^\dagger_{i,\alpha} $ and $a_{i,j} \to a_{i,j} - \theta_i +\theta_j$.
The \textit{bonafide} DSL wave function is constructed by applying the Gutzwiller projector ${\mathcal{P}_G=\prod_i (n_{i,\uparrow}-n_{i,\downarrow})^2}$ (with ${n_{i,\alpha}=f^\dagger_{i,\alpha} f^\dagga_{i,\alpha}}$) to the ground state of $\mathcal{H}_0$. This state, which has no free variational parameters, turns out to be remarkably stable to symmetry breaking and lowering of the U(1) gauge structure to $\mathbb{Z}_2$~\cite{iqbal2016}.
In the low-energy (IR) limit, this is a theory of $N_f=4$ Dirac cones coupled to a U(1) gauge field, or QED$_3$, which is believed to flow to a conformal fixed point \cite{PhysRevD.105.085008}.
Importantly, there are gapless excitations of the (compact) gauge field as monopoles $\Phi$, i.e. tunneling events of the emergent magnetic flux. These are highly relevant operators about the conformal fixed point, estimated to have a scaling dimension $\Delta_\Phi$ within the range $1.18$--$1.34$ \cite{PhysRevD.100.054514}.

\begin{figure}
\includegraphics[width=\columnwidth]{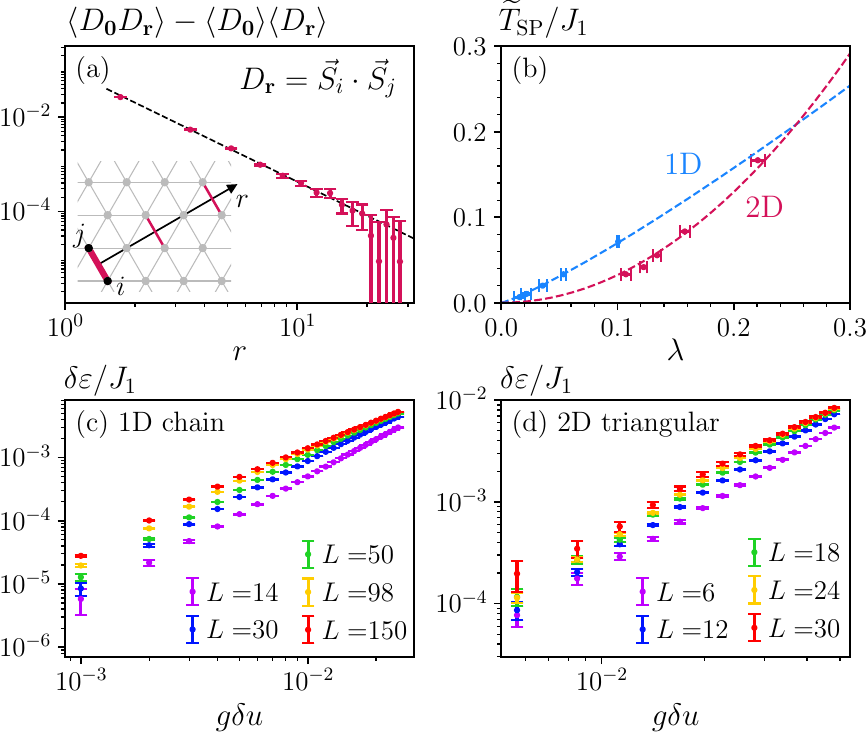}
\caption{\label{fig:static} Static limit of lattice distortions and weak coupling physics. (a) Dimer-dimer correlations of the triangular-lattice DSL, fitted with a power-law $r^{-2\Delta_\Phi}$, $\Delta_\Phi=1.20(3)$ (b) Spin-Peierls temperature $\widetilde{T}_{\mathrm{SP}}$ as a function of the dimensionless spin-lattice coupling $\lambda$, for the 1D antiferromagnetic Heisenberg chain and the DSL ansatz for the triangular-lattice $J_1$-$J_2$ model (2D). See Supplementary Material for details of the fitted curves, obtained from the data collapse of Fig.~\ref{fig:collapses}. (c,d) The energy gain per site $\delta\varepsilon$, as a function of static distortions, for various linear sizes $L$ in 1D and 2D (where $N=L\times L$ sites).}
\end{figure}

Theoretical works predict that a subset of the monopole excitations of the DSL have zero spin and transform as VBS order parameters with momentum $\bm{X}=\bm{K}/2$, where $\bm{K}$ is the corner of the Brillouin zone~\cite{song2019,song2020}, as seemingly confirmed by numerical results of the $J_1$--$J_2$ model on small clusters~\cite{wietek2023}. 
Motivated by these findings, we look at the dimer-dimer correlations $\mathcal{C}(r)$ of the triangular-lattice DSL, shown in Fig.~\ref{fig:static}(a) as a function of the spatial separation $r$.
We find a power-law decay $\mathcal{C}(\bm{r}) \propto r^{-2\Delta_\Phi}$, where $\Delta_\Phi=1.20(3)$, and a spatial pattern consistent with the wavevector $\bm{X}$. 
This is compatible with a scenario where the dimer-dimer correlations are dominated by spin-singlet monopoles \cite{song2019,song2020,wietek2023}, and we hence interpret $\Delta_\Phi$ as the monopole scaling dimension.

\paragraph{Static distortions.}
In this section we evaluate the response of the DSL to static, longitudinal, spatially uniform lattice distortions with momentum $\bm{X}$ commensurate with the monopoles. A consequence of their power-law correlations is the potential divergence of the system's susceptibility and a corresponding instability~\cite{seifert2024}.

The distortions are coupled to the spins by assuming, without loss of generality, that the first-neighbor exchange interactions depend linearly on the relative sites displacements as follows
\begin{align}\label{eq:static_ham}
{\cal H}[\bm{u}_i] = {\cal H}_{J_1 J_2} &-gJ_1 \sum_{\langle i,j\rangle}  \,\hat{\bm{r}}_{i,j} \cdot(\bm{u}_i-\bm{u}_j)  \vec{S}_i \cdot \vec{S}_{j} \, ,
\end{align}
with $g$ denoting the spin-lattice coupling constant and $\bm{u}_i$ the local displacement operators.
The unit vectors $\hat{\bm{r}}_{i,j}$ are defined as $\hat{\bm{r}}_{i,j}=\frac{\bm{r}_i-\bm{r}_j}{\| \bm{r}_i-\bm{r}_j\|}$, with $\{\bm{r}_i\}$ indicating the site coordinates in the undistorted triangular lattice. 
This form of the spin-phonon interaction is a strong-coupling analog of the Su--Schrieffer--Heeger model for electron-phonon systems~\cite{su1979}.
In this static limit, the lattice dynamics is frozen and the displacement operators $\bm{u}_i$ become classical variables with energy cost $\frac{1}{2} K\, \bm{u}_i^{\ 2} $ per site.
The distortion we impose is a 12-site pattern of strength $\delta u$, commensurate with the monopole momenta~\cite{seifert2024} (see Supplementary Material).
In parallel, we repeat the same analysis in the 1D Heisenberg chain, where the lattice distortion is a dimerization ${u_i = \delta u (-1)^i}$. 
We normalize both cases such that the energy cost (per site) of lattice displacements is $\frac{1}{2} K \, \delta u^2$ and hence the dimensionless spin-lattice coupling is $\lambda = J_1\, g^2 /K$.

We compute how the variational ground state energy of the $J_1$--$J_2$ model in Eq.~\eqref{eq:static_ham} changes when the exchange couplings are modulated by lattice distortions, on a set of $L\times L$ geometries.
The energy gain per site $\delta \varepsilon$ (w.r.t. the undistorted case $\delta u=0$) is shown in Fig.~\ref{fig:static}(d), as a function of the distortion strength and for various linear system sizes $L$. 
The same calculation of the 1D Heisenberg chain's energetic response to dimerization as a function of length $L$ is shown in Fig.~\ref{fig:static}(c).
For small distortions, we observe a quadratic gain in energy $\delta \varepsilon = A(L) g^2\, \delta u^2$ that is strongly $L$-dependent.
This energy gain competes with the quadratic energy cost of static distortions to give the effective potential $\epsilon (\delta u) = (K/2)(\delta u)^2 - \delta \epsilon$.
Thus, the lattice will spontaneously distort whenever the sign of this quadratic response changes to negative, i.e. when $2A(L_c) = K/g^2$, yielding a critical system size $L_c$ above which the lattice will spontaneously distort.
Because algebraic QSLs are relativistic theories with dynamical scaling exponent $z=1$, $L_c$ can be understood to define an effective spin-Peierls temperature scale $\T_{\mathrm{SP}} / J_1 = L_c^{\ -1}$, below which an infinitely large system will distort.
This is motivated by the observation that CFTs at finite temperature are described by a finite-size ($\beta = T^{-1}$) in the Euclidean time direction \footnote{In this construction, the effective temperature $\tilde{T}$ is related to the true temperature $T$ by a (geometry-dependent) $\mathcal{O}(1)$ constant factor in 2D. Note that in 1D the effective critical temperature is exactly the spin-Peierls temperature, see Supplementary Material for details.}.
We estimate the spin-Peierls transition line by measuring the response $A(\T)$ for various effective temperatures; then, the critical (dimensionless) spin-lattice coupling strength is $\lambda_c(\T) = J_1 g_c^2/K = J_1/[2 A(\T)]$.
For a given coupling strength, this equivalently defines an effective transition temperature $\T_{\mathrm{SP}}$ which is plotted as points in Figure~\ref{fig:static}(b) for 1D and 2D.

\paragraph{Scaling collapse.}
We will now discuss in detail the possibility that this spin-Peierls transition is precipitated by lattice couplings to emergent gapless monopoles and argue for the presence of a weak-coupling instability.
To do this, we highlight that our numerical results on larger systems show significant deviation from a quadratic (weak-coupling) response: the energy gain becomes significantly less $L$-dependent and is best described by a lower exponent $(g \delta u)^\chi$, $\chi<2$.
In the thermodynamic limit, we predict that the static distortion will open a gap by inducing a relevant instanton; the respective field-theoretic coupling has scaling dimensions $d-\Delta$, where $d=D+1$ and we define $\Delta=\Delta_*,\Delta_\Phi$ in 1D and 2D, respectively.
In our calculations, the system size also acts as a relevant operator with the coupling $1/L$ (scaling dimension $1$) which opens a finite-size gap \cite{Cardy_1996}.
To this end, we make the following scaling Ansatz for the energy density gain
\begin{equation}\label{eq:energy_collapse}
    \delta\varepsilon(g\, \delta u, 1/L)/J_1 = L^{-d} \tilde{\mathcal{F}}\left(L^{d-\Delta} \, g\, \delta u\right),
\end{equation}
in terms of a universal scaling function $\tilde{\mathcal{F}}(x)$. 
In the weak-coupling (small distortions, small system) limit, i.e. ${x\to 0}$, the function must be quadratic $\tilde{\mathcal{F}}(x)\sim x^2$; conversely in the large-parameter, \emph{strong-coupling}, limit ${x\to\infty}$, the function must follow $\tilde{\mathcal{F}}(x)\sim x^{d/(d-\Delta)}$ in order to be $L$-independent, thereby reproducing non-analytical scaling laws determined by the CFT fixed point theory.

\begin{figure}
\includegraphics[width=\columnwidth]{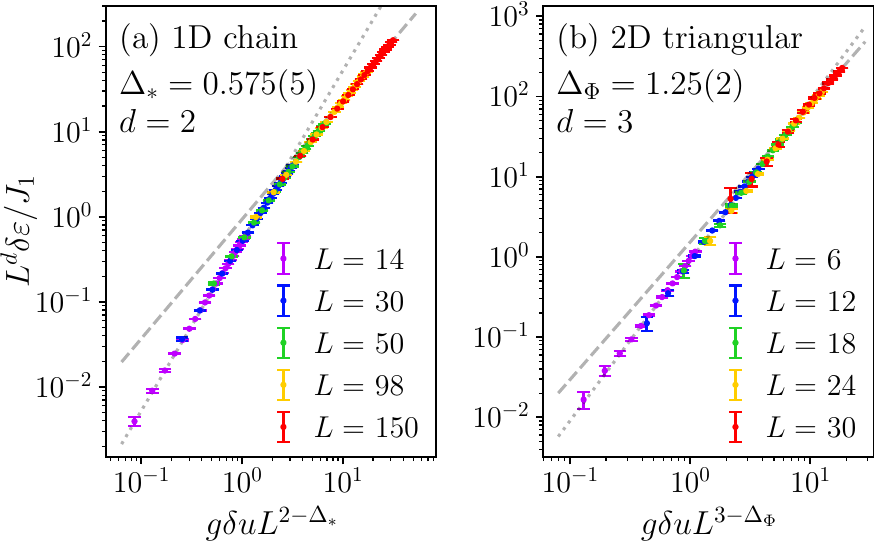}
\caption{\label{fig:collapses} Data collapse of the energy gain per site $\delta \varepsilon$ under static lattice distortion $g\, \delta u$. Results for (a) the Heisenberg chain and (b) the triangular lattice $J_1$--$J_2$ model are shown. 
We plot $L^d\delta\varepsilon / J_1$ against $g\, \delta u \, L^{d-\Delta}$ for dimension $d=D+1$ and optimise the scaling dimensions to get the best collapse, shown inset.
Dotted and dashed line represent asymptotic fitting functions for the regimes of weak and strong coupling, i.e. at small and large values of $g\, \delta u$. 
}
\end{figure}

%
We apply this analysis to the $d=2+1$ U(1) DSL in Fig.~\ref{fig:collapses}(b) and extract the exponent $\Delta_\Phi = 1.25(2)$. This agrees remarkably well with the fit of dimer-dimer correlations and is compatible with independent predictions of the QED$_3$ monopole scaling dimension \cite{PhysRevD.100.054514}.
A power-law fit to the weak- and strong-coupling regimes are shown as dotted and dashed lines in Fig.~\ref{fig:collapses}(b).
The weak-coupling fit allows us to obtain the functional form of the critical temperature as a function of coupling strength by using the same procedure as for individual data points above. We find in 2D $\T_{\mathrm{SP}} / J_1 \approx 3.24 \, \lambda^{2.0}$, as plotted with the dashed line in Fig.~\ref{fig:static}(b). 
This static calculation lies on the $\omega=0$ axis of the phase diagram Fig.~\ref{fig:phasediag}.

We repeat the analysis for the $d=1+1$ Heisenberg chain data in Fig.~\ref{fig:collapses}(a).
The extracted exponent is $\Delta_* = 0.575$ and we measure the spin-Peierls temperature $\T_{\mathrm{SP}} / J_1 \approx 1.05 \, \lambda^{1.18}$; the exponent is shifted from the bare scaling dimension $0.5$ and the transition temperature is suppressed compared to the approximate field theoretic prediction due to logarithmic corrections \footnote{This effect is described in previous works \cite{papenbrock2003,orignac2004a}, see Supplementary Material for a discussion.}.
Taken together, we find similar behaviour of static 1D and 2D spin-Peierls transitions, suggesting that for comparable spin-lattice couplings, the spin-Peierls transition in two dimensions could be an equally important instability as in 1D.

In both models, the strong-coupling regimes are also well fitted by the proposed power-laws [with fixed exponent $d/(d-\Delta) < 2$].
This implies the spin energy gain will always outcompete the lattice stiffness in the thermodynamic limit, such that the \textit{static} model has a gapped ground state for any non-zero spin-lattice coupling.
The magnitude of the strong-coupling response controls the size of the equilibrium lattice distortion and spin-gap, which may be potentially measured in experiments.

\paragraph{Dynamical phonons.}
We will now study a full model in which lattice distortions (i.e., phonons) are treated quantum mechanically, allowing us to calculate the phase diagram of the spin-Peierls transition as a function of phonon frequency $\omega$ and assess how the static instability is dampened or resolved.
We focus here on the U(1) DSL on the triangular lattice, by considering the Hamiltonian
\begin{align}
{\cal H} = {\cal H}[\bm{u}_i] +
 \sum_i \left (\frac{\bm{p}_i^{\ 2}}{2m} + \frac{1}{2} m \omega^2 \, \bm{u}_i^{\ 2} \right). \label{eq:spinphonon}
\end{align}
where $\bm{u}_i$ and $\bm{p}_i$ are displacement and momentum operators of uncoupled harmonic oscillators (i.e. Einstein phonons); these are coupled to the spins by Eq.~\eqref{eq:static_ham}.
We pursue a full quantum treatment of the Hamiltonian~\eqref{eq:spinphonon}, by means of a recently developed VMC approach. The variational Ans\"atze take the form $|\Psi_{\mathrm{sp}}\rangle=\mathcal{J}_{\mathrm{sp}} |\Psi_{\mathrm{p}}\rangle \otimes |\Psi_{\mathrm{s}}\rangle$, where 
$|\Psi_{\mathrm{s}}\rangle$ is a Gutzwiller-projected fermionic state, $|\Psi_{\mathrm{p}}\rangle$ is a phonon coherent state describing lattice distortions, and $\mathcal{J}_{\mathrm{sp}}$ is a spin-phonon Jastrow factor~\cite{ferrari2020a,ferrari2021,ferrari2024}. 
At variance with the case of static distortions, we do not impose a specific phonon mode, but we instead perform unbiased optimizations of the variational state to find the lowest-energy distortion.
Details about the VMC method are provided in the Supplementary Material. 
At large enough spin-phonon coupling ($g>g_c$), our variational search finds that the system forms a 12-site VBS state with the distortion pattern shown in the inset of Fig.~\ref{fig:phasediag}, compatible with field theory predictions.
We then estimate the critical coupling $g_c(\omega)$ at which phonons condense and distort the lattice, by comparing the energies of the VBS and the (undistorted) DSL Ans\"atze.

\begin{figure}[t]
\includegraphics[width=0.48\textwidth]{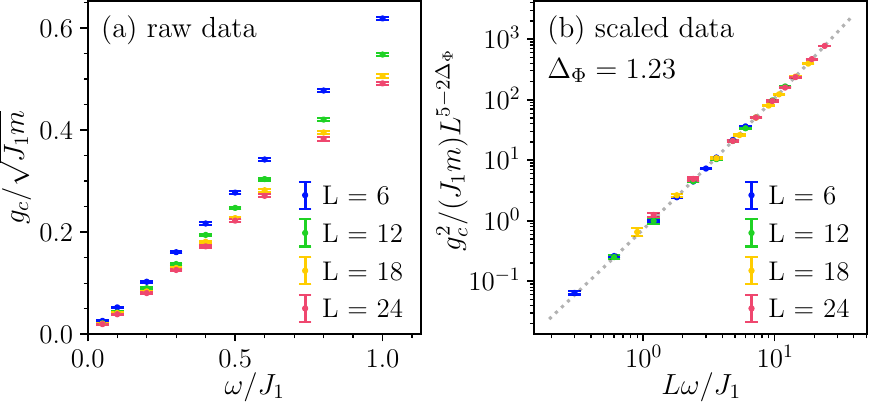}
\caption{\label{fig:g_scaling} (a) Measured critical phonon coupling strength $g_c$ of the U(1) DSL as a function of phonon frequency $\omega$ (dynamical phonons) and linear system size $L$. 
We produce a data collapse in (b) by plotting $g_c^2/(J_1 m) L^{5-2\Delta_\Phi}$ against $L\, \omega/J_1$. }
\end{figure}

The results are shown for varying system sizes $L$ in Fig.~\ref{fig:g_scaling}(a), where the critical coupling $g_c$ is plotted (in units of $\sqrt{J_1 m}$).
Firstly, we see that as we take the adiabatic limit $\omega\to 0$ the critical coupling approaches zero linearly (with an $L$-dependent coefficient), in agreement with our static results that find an instability, i.e. $g_c(\omega=0)=0$.
We observe a deviation from the linear trend for larger $\omega$; in order to capture the system-size dependence in this strong-coupling regime, we employ another scaling Ansatz.
In the field theoretical description of the model, dynamical phonons can be integrated-out exactly to leave an effective retarded interaction between monopoles, with an effective coupling constant $\eta = g^2/(J_1 m)$
with field-theoretic scaling dimension $d+2-2\Delta_\Phi$.
Our Ansatz for the critical coupling $\eta_c$ as a function of $\omega/J_1$ and $1/L$ reads
\begin{equation}
    \eta_c(\omega/J_1, 1/L) = g^2_c/(J_1 m)
    = L^{2\Delta_\Phi - d - 2}\, \tilde{\mathcal{G}}(L \, \omega / J_1)
\end{equation}
and the collapse is achieved by plotting $\eta_c \, L^{d+2-2\Delta_\Phi}$
against $L \, \omega/J_1$. In the 2D case considered here, we find the best collapse for $\Delta_\Phi=1.23(3)$, as shown in Fig.~\ref{fig:g_scaling}(b). 
The scaling function must be quadratic in the small-parameter regime; by fitting the collapsed data to a quadratic function $\tilde{\mathcal{G}}(x) = a_1\, x^{2}$, we extract the critical (effective) temperature as a function of the spin-lattice coupling as $\omega\to 0$.
Explicitly, this is done by using the scaling $\eta_c = a_1(\T/J_1)^{d-2\Delta_\Phi} (\omega/J_1)^2$
and then 
evaluating the dimensionless (critical) coupling $\lambda_c = J_1 \, g^2 / (J_1 m) = a_1(\T/J_1)^{d-2\Delta_\Phi}$. Here, we used $K=m\omega^2$ and took $\omega\to 0$ with $K$ fixed.
The resulting function $\T_{\mathrm{SP}} / J_1 \approx 2.72 \, \lambda^{1.85}$ agrees numerically with the independent static calculation from Fig.~\ref{fig:static}(b) (within error bars for small $\lambda$).
In the large-parameter regime, we similarly extract the critical frequency $\omega_c / J_1 \approx 39\,\lambda^{1.85}$ below which the $\T=0$ ground state is spin-Peierls ordered (equivalently the thermodynamic limit in our calculations).
In Fig.~\ref{fig:phasediag} these critical points $\T_{\mathrm{SP}}$ and $\omega_c$ are plotted for $\lambda=0.3$.
The phase boundary $\T_c(\omega)$ for $0 < \omega <\omega_c$ is plotted using a fit to the whole scaling function which interpolates the two regimes, shown as a dotted line in Fig.~\ref{fig:g_scaling} and discussed in the Supplementary Material.

\paragraph{Outlook.}
Using large scale VMC simulations for the $J_1$--$J_2$ triangular lattice Heisenberg model, we report a power-law decay of the dimer-dimer correlation function. This is consistent with predictions of the conformal continuum theory of the U(1) DSL.
By studying this model coupled to phonons, both in the static limit and in the full quantum regime, we confirm that the spin-Peierls instability is relevant in 2D, but with transition temperatures that are suppressed in the regime of large phonon frequencies, 
as is potentially significant for current experiments with $J_1$--$J_2$ spin-liquid candidate materials \cite{scheie2024,scheie2024b}.
Recently, a very low-energy gap has been observed in NaYbSe$_2$ which could be related to such a weak spin-Peierls transition.
Thus, it is important to search for signs of lattice symmetry breaking compatible with the VBS pattern of our spin-Peierls scenario, an observation of which would serve as a distinctive signature of an underlying U(1) DSL state. 

There exist other exotic deconfined phases of emergent U(1) gauge theories in two dimensions such as deconfined quantum critical points, as potentially realized in $J$--$Q$ type models \cite{Senthil2004,Senthil2004B,sandvik2010}. It is an important question to understand the stability of these transitions to spin-lattice coupling or alternatively the possibility that they might be driven by varying the phonon energy scale \cite{hofmeier2024,gotz2024}.
Our work demonstrates that there exists a strong similarity of spin-Peierls instabilities of algebraic spin liquids in 1D and 2D, both in the static case and upon coupling to quantum phonons. It is an interesting direction for future studies to investigate to what extent such analogies are also present for other deformations of algebraic spin liquids in 1D and 2D, e.g. interchain/interlayer couplings and disorder, which may be controlled by monopoles or potentially other exotic quasiparticles in the spectrum.

\vspace{5mm}

\paragraph{Acknowledgments.}
We thank F. Becca, M. Drescher and F. Pollmann for insightful discussions.
We acknowledge support by the Deutsche Forschungsgemeinschaft (DFG, German Research Foundation) 
for funding through TRR 288 -- 422213477, project A05 (F.F. and R.V.), through Project No. 509751747 (VA 117/23-1) (R.V.), through a Walter Benjamin fellowship, Project ID 449890867 (U.F.P.S), and through TRR 360 (Project-ID No. 492547816) (J.W. and J.K.). 
This research was supported in part by the National Science Foundation under Grant No. NSF PHY-1748958.
J.K. acknowledges support from the Imperial-Technical University of Munich flagship partnership.
The research is part of the Munich Quantum Valley, which is supported by the Bavarian state government with funds from the Hightech Agenda Bayern Plus. 
This work was performed in part at Aspen Center for Physics, which is supported by National Science Foundation grant PHY-2210452. 

\bibliography{main}

\clearpage

\section{Supplementary Material}

In this Supplementary Material we present the computational details of the VMC calculations performed in this work and the field theory describing the spin-lattice coupling, which guides the analysis of the numerical data. 
The aim of the VMC simulations is understanding how the interaction between spins and lattice distortions [Eq.~\eqref{eq:static_ham}] affects the DSL ground state of the triangular-lattice $J_1$--$J_2$ model. Analogous calculations are performed in 1D for the nearest-neighbor Heisenberg chain. On the one hand we consider the limit of static (i.e., classical) distortions, which  boils down to the study of a spin model with bond-dependent exchange couplings. On the other hand, we perform simulations of the full spin-lattice problem, treating both spins and distortions (i.e., phonons) at the quantum level.
We also briefly outline the field theoretical descriptions of the 1 and 2D models coupled to lattice distortions, and use these to derive the form of the scaling collapse which we use in the analysis of the numerical data.

\subsection{Spin models}

Both spin models under investigation can be described by the Hamiltonian of Eq.~\eqref{eq:spin_hamiltonian}. We study the the $J_1$--$J_2$ model on the 2D triangular lattice (at $J_2/J_1=1/8$), and the 1D Heisenberg model on a linear chain ($J_2=0$). The VMC simulations are performed on finite-size clusters with periodic boundary conditions. The triangular-lattice clusters contain $L^2$ sites and are defined by the translation vectors $L \bm{a}_1$ and $L \bm{a}_2$, with $\bm{a}_1=(1,0)$ and $\bm{a}_2=(\frac{1}{2},\frac{\sqrt{3}}{2})$. In 1D we consider linear chains of $L$ sites.

\subsubsection{Gutzwiller-projected states}

The variational wave functions (for spins) are projected fermionic states of the form $|\Psi_s\rangle=\mathcal{P}_G |\Phi_0\rangle$, in which the Gutzwiller projector ${\mathcal{P}_G=\prod_i (n_{i,\uparrow}-n_{i,\downarrow})^2}$ is applied to a wave functions of Abrikosov fermions $|\Phi_0\rangle$~\cite{wen2002,beccabook}. The latter is defined as the ground state of an auxiliary fermionic Hamiltonian $\mathcal{H}_0$, which in general contains real-valued hoppings $t_{i,j}$ and singlet pairings $\Delta_{i,j}=\Delta_{j,i}$ (variational parameters)
\begin{equation}
{\cal H}_{0} = \sum_{i,j} \sum_{\alpha=\uparrow,\downarrow} t_{i,j} f_{i,\alpha}^\dagger f_{j,\alpha}^\dagga +
\sum_{i,j}  \Delta_{i,j} f_{i,\downarrow}^\dagga f_{j,\uparrow}^\dagga + h.c. 
\label{eq:aux_H0}
\end{equation}
For the triangular-lattice $J_1$--$J_2$ model at ${J_2/J_1=1/8}$ the optimal ground state is the U(1) DSL~\cite{iqbal2016}. In this case ${\cal H}_{0}$ contains only a nearest-neighbor hopping, with uniform absolute value and a bond-dependent sign, which creates a pattern of alternating $0$ and $\pi$ fluxes in the triangular plaquettes (see inset of Fig.~\ref{fig:phasediag}). The spectrum of ${\cal H}_{0}$ is gapless with two Dirac points (for each spin component).
We note that although the debate concerning the nature of the $J_1$--$J_2$ model's QSL ground state and excitations is not fully settled, recent numerical results corroborate the DSL scenario \cite{hu2019,sherman2023,drescher2023,wietek2023}.
For the 1D Heisenberg chain, the lowest variational energy is achieved by taking (translationally invariant) hopping and pairing at first-neighbor, plus an additional hopping at third-neighbors, within $\mathcal{H}_0$~\cite{becca2011,ferrari2018}. The resulting fermionic band of  $\mathcal{H}_0$ is gapless at $k=\pm \frac{\pi}{2}$. 
The optimal values of the variational parameters can be obtained by a numerical minimization of the variational energy $E_{\rm var}$. For this purpose, in all VMC calculations performed in this work, including the ones involving phonons, we use the stochastic reconfiguration method~\cite{sorella2005,beccabook}.

\subsection{Static distortions}

\begin{figure*}
\includegraphics[width=0.85\textwidth]{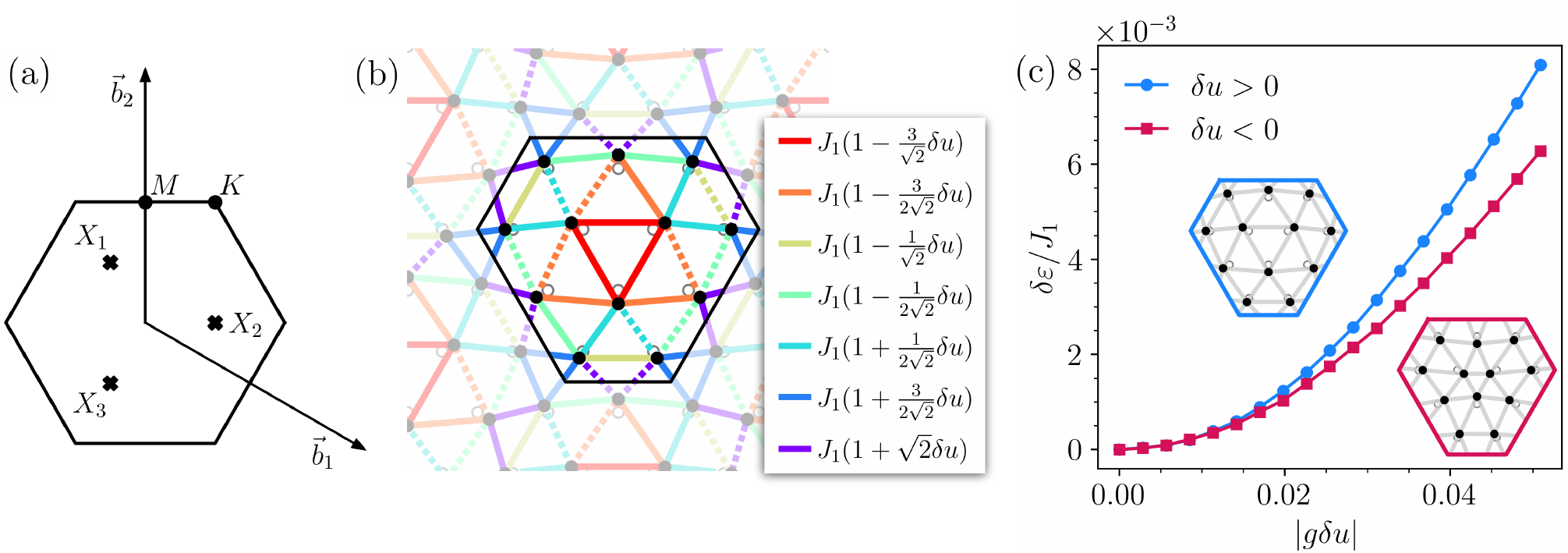}
\caption{Panel (a): Brillouin zone of the triangular lattice with reciprocal lattice vectors and high symmetry points. Panel (b): lattice distortion giving the strongest response on the triangular lattice. Lattice sites are displaced according to Eq.~\eqref{eq:dist_2d} (with $\delta u >0$). The resulting 12-site supercell is delimited by the solid black lines. Different colors represent symmetry-inequivalent first-neighbor bonds, grouped into distinct classes according to their bond length after distortion. For each class of bonds, the value of the exchange interaction in presence of a finite static distortion $\delta u$ is reported. The variational parametrization of $\mathcal{H}_0$ follows the same symmetry of the distortions: hoppings with different absolute values are taken on the different classes of inequivalent bonds. The signs of the hoppings, instead, are kept equal to the ones of the DSL and are represented here by solid/dashed lines, which indicate positive/negative hoppings, respectively. Panel (c):
energy gain per site $
\delta \varepsilon$ as a function of the lattice distortions of Eq.~\eqref{eq:dist_2d}, for different signs of $\delta u$. Results for the $L=12$ lattice. The respective sites displacements are shown in the insets. The case $\delta u>0$ provides a larger energy gain beyond the weak-coupling regime of small distortions.  \label{fig:supp_class}
}
\end{figure*}

\subsubsection{Lattice distortions in the static limit}

The effect of the spin-lattice interaction is modeled by a linear coupling between the (first-neighbor) exchange interaction and the relative sites displacements, as in Eq.~\eqref{eq:static_ham}. We note that the sign in front of the spin-phonon coupling parameter $g$ is in principle arbitrary, as the model is invariant under $\bm{u}_i \mapsto -\bm{u}_i$ and $g \mapsto -g$. We choose to include a minus sign in Eq.~\eqref{eq:static_ham} such that, for $g>0$, stronger spin-spin correlations form on shorter bonds, as expected from physical grounds; in other words, the antiferromagnetic exchange interaction is enhanced when spins get closer. 

In the limit of \textit{static distortions}, the displacements $\bm{u}_i$ are classical variables and they induce a static modulation of the exchange couplings. We can indeed fix the values of the sites displacements according to certain phonon modes and obtain a spin model with bond-dependent exchange couplings. We then compute the variational energy gain (per site) $\delta \varepsilon$ as a function of the strength of lattice distortions.

Let us consider the simpler 1D model as the showcase example. The pattern of distortions which provides the larger energy gain is the staggered dimerization ${u_i= (-1)^i \delta u}$~\cite{cross1979}. When $\delta u$ is finite, the exchange interactions become $J_1(1\pm 2g\delta u)$ on even/odd bonds, respectively. We thus compute the ground state energy of the spin model with alternating exchange interactions for different values of $\delta u$. The energy gain per site is then given by $\delta \varepsilon = |E_{\rm var}(\delta u)-E_{\rm var}(0)|/N$ (with $N=L$ in 1D). Since the distortions lower the translational symmetry of the model and give rise to a doubled unit cell, we increase the variational freedom of our state accordingly, i.e. we allow the hoppings and pairings of $\mathcal{H}_0$ to take different values on even and odd bonds in the numerical optimization. This yields a gapped dimerized state at finite $\delta u$.

An analogous approach is followed for the 2D triangular lattice model. In this case, the choice of the distortion pattern is more complicated given the larger dimensionality. Theoretical predictions and numerical calculations in~\cite{seifert2024} showed that the strongest response is given by the distortion 
\begin{align}
    \bm{u}_i= \sqrt{2}\, \delta u \Im
    \left[\sum_\alpha s_\alpha e^{i\bm{X}_\alpha \cdot \bm{r}_i } \frac{\bm{X}_\alpha}{\| \bm{X}_\alpha \|}\right]\, , \label{eq:dist_2d}\\
    \text{with  } s_\alpha=\frac{1}{\sqrt{3}}(-1,-e^{i 2\pi/3},e^{-i 2\pi/3}) \, .
    \label{eq:dist_2d_sa}
\end{align}
This corresponds to a $D_3$-symmetric combination of longitudinal phonon modes with lattice momentum $\bm{X}_\alpha$ [$\alpha=1,2,3$, see Fig.~\ref{fig:supp_class}(a)], compatible with the momenta of the singlet monopole operators of the DSL. The pattern of displacements is described by a 12-site supercell [defined by the vectors $2( \bm{a}_1 \pm \bm{a}_2)$], which contains symmetry-inequivalent first-neighbor bonds that can be grouped in 7 classes, according to their length after distortion, as depicted in Fig.~\ref{fig:supp_class}(b). Therefore, we let the variational hopping parameters of $\mathcal{H}_0$ take independent values on the symmetry-inequivalent bonds. Specifically, the absolute values of the inequivalent hoppings are independently optimized, while the underlying sign structure is kept fixed and equal to the one of the DSL. Analogously to the 1D case, the spectrum of the optimal fermionic Hamiltonian $\mathcal{H}_0$ acquires a gap in presence of finite distortions $\delta u$. To verify the validity of our parametrization, we performed also a full unbiased optimization of all the 36 hoppings within the 12-site supercell, which returned the same variational energy.

\subsubsection{Static distortions: Results}

We cross-checked that the energy gain provided by the displacements of Eq.\eqref{eq:dist_2d} is larger than the one obtained by the $\bm{M}$ and $\bm{K}$ distortions considered in \cite{seifert2024}. 
We note however that, contrary to the 1D case, the sign of $\delta u$ in Eq.~\eqref{eq:dist_2d} is relevant, as it gives rise to inequivalent patterns of distortions. Such sign cannot be determined by the field theory analysis of~\cite{seifert2024} (or their weak-coupling numerical results) and may only be determined numerically on larger systems.
We find that $\delta u>0$ [in the conventions of \eqref{eq:dist_2d} and~\eqref{eq:dist_2d_sa}] provides a larger energy gain beyond the weak-coupling regime, as shown in Fig.~\ref{fig:supp_class}(c).
Furthermore, the resulting distortion pattern for $\delta u>0$ is compatible with the one found by the calculations with dynamical phonons [see Fig.~\eqref{fig:phasediag}], where it emerges spontaneously from an unbiased optimization of the phonon variational parameters (see discussion below). 

We then analyzed the system-size dependence of the responses ($\delta \varepsilon $ vs $g\delta u$) of both the 1 and 2D systems to verify the presence of an instability in the thermodynamic limit. This was done by first fitting each curve of Figs.~\ref{fig:static}(c,d) to a quadratic function in the small-$g\delta u$ regime and extracting the coefficients $A(L)$. We then interpret a critical temperature using the method described in the main text.
The range of data to fit to a quadratic are chosen for each system individually, such that the quadratic function produces a reasonable fit across the full range. We account for the potential systematic error introduced by this method in two ways.
Firstly, we extract the errors of the amplitudes [and plot the resulting error of the critical temperatures in Fig.~\ref{fig:static}(b)] by evaluating how far the first excluded data point lies from the fit curve. Secondly, we turn to an alternative method of fitting to produce the continuous curves of Fig.~\ref{fig:static}(b); this method is detailed in the following section of this Supplementary Material and inspired by the field-theoretic descriptions of the respective spin liquids. The advantage of this approach is that it does not rely on fitting the individual data sets (which may have as many as 16 but as few as 3 points in the weak-coupling regime), but uses a collapse to transform them into one. Both sets of collapsed data can then be fit to a quadratic function for weak coupling (small rescaled $x$ parameter) to produce the critical temperature curves.

\subsection{Data collapse}

Here we briefly develop 2D QED$_3$ and the 1D Luttinger liquid as continuum field theories in order to describe their coupling to the lattice. These theories capture the low energy behaviour of the 1D and 2D antiferromagnetic Heisenberg models considered in this work.

\subsubsection{Lattice coupling: QED$_3$}
In the continuum description of the U(1) DSL with $N_f=4$ fermions, the most relevant spin-singlet operators are the (three degenerate) monopoles $\Phi_\alpha(x)$.
In the leading-order large-$N_f$ expansion, the monopole scaling dimension is $\Delta_\Phi=1.02$.
This is in agreement with conformal bootstrap bounds (when one assumes the DSL is stable) \cite{10.21468/SciPostPhys.13.2.014,PhysRevD.105.085008}, and numerical simulations of lattice QED$_3$ put $\Delta_\Phi$ within the range $1.18$--$1.34$ \cite{PhysRevD.100.054514}
In the triangular lattice model, these transform as VBS order parameters $\vec{S}_i\cdot \vec{S}_{j}\sim \Re \Phi_\alpha(x) \, e^{i \bm{X}_\alpha.\bm{r}_i}$ with a finite momentum eigenvalue $\bm{X}_\alpha = -\bm{K}_\alpha/2$ with $\alpha=1,2,3$ \cite{song2019,song2020} ($\alpha=1,2,3$ is a valley index). This identification must lead to algebraic decay of dimer-dimer correlation functions at this momentum, with an exponent fixed by the monopole CFT scaling dimension $\mathcal{C}(r)\sim r^{-2\Delta_\Phi}$. 
Given their nontrivial transformation properties, the lowest-lying symmetry-allowed operator is the triple ($6\pi$-flux) VBS-monopole $\Phi_1\Phi_2\Phi_3$, which is believed to be irrelevant \cite{chester16}, meaning the theory may flow to a conformally-invariant fixed point.

A classical distortion field, as given in Eq.~\eqref{eq:dist_2d}, perturbs the action $S_{\mathrm{QED}_3}$ by introducing a coupling between the spin-singlet monopoles and longitudinal distortion modes \cite{seifert2024}. We write the overall spin action as
\begin{multline}\label{eq:qed3_lattice_coupling}
    S[\delta u] = S_{\mathrm{QED}_3} + \tilde{g} \, \delta u\int\dd[3]{x} \left[\mathcal{V}^\dagger(x) + \mathrm{h.c.}\right] \, , \\
    \mathcal{V}^\dagger = -\sum_\alpha s_\alpha\Phi^\dagger_\alpha
    = \Phi^\dagger_1 + e^{\pi i / 3} \Phi_2^\dagger + e^{2\pi i / 3} \Phi_3^\dagger \, ,
\end{multline}
where $\delta u$ is the strength of the distortion and we inserted the phases $s_\alpha$ from \eqref{eq:dist_2d_sa}. 
The operator $\mathcal{V}^\dagger$ proliferates 12-site VBS order \cite{song2019}, and its field-theoretic lattice coupling $\tilde{g}$ is non-trivially related to the microscopic spin-lattice coupling $g$ defined in \eqref{eq:static_ham} of the main text.

\subsubsection{Lattice coupling: Luttinger liquid}

The spin-liquid ground state of the 1D Heisenberg chain is understood in terms of a similar continuum field theory:
This model is described by the Sine--Gordon action $S_{\mathrm{SG}}[\phi]$, the effective theory of the Luttinger liquid in terms of a bosonic field $\phi(x)$. The local order parameters may be identified $\vec{S}_i = \cos[2\phi(x_i)]$ and $D_i = \vec{S}_i\cdot\vec{S}_{i+1} = \sin[2\phi(x_i)]$. These have scaling dimension $\Delta=1/2$, in the SU(2) symmetric (nearest-neighbor) Heisenberg model, meaning correlation dimer-dimer correlation functions should decay as $\mathcal{C}(r) = r^{-2\Delta}$, with $2\Delta = 1$.
However, in this model the most relevant symmetry-allowed operators are $\cos[4\phi(x_i)]$, $\sin[4\phi(x_i)]$, which are marginal $4\Delta=2$.
Beyond the classical scaling dimension, these operators are marginally irrelevant, allowing the SU(2) symmetric model to be a stable gapless ground state, but the coupling of the renormalization group equations between $\sin[4\phi(x_i)]$ and $\sin[2\phi(x_i)]$ operators leads to logarithmic corrections to the correlation functions
At long distances, this leads to an effective shift of the correlation function decay exponent $2\Delta\to 2\Delta_* = 1+A/2$ ($A>0$). In practice, this will mean that correlation functions (and other observables which may be written in terms of it) will be controlled by an effective scaling dimension $\Delta_* > 1/2$.

In 1D, the spin-lattice coupling is written as follows~\cite{orignac2004}
\begin{equation}
        S[\delta u] = S_{\mathrm{SG}} + \tilde{g} \, \delta u \int \dd{x}\dd{\tau} \sin(2\phi)\,.
    \label{eq:SG_lattice}
\end{equation}
In this case, we can utilize the bosonization mapping to relate field-theoretic couplings to microscopic parameters
\begin{equation}
    \tilde{g} = \frac{6 J_1}{a\pi^2}\left(\frac{\pi}{4}\right)^{1/4}g\, ,
\end{equation}
following Ref.~\cite{orignac2004} in our conventions.

\subsubsection{Weak-coupling response: Finite temperature versus finite systems}
We now attempt to describe the physics around the spin-Peierls transition; here, the order parameter will be small and so we can expand in small $\delta u$ to find its effective (quadratic) potential. When this inverts, there will be an instability of the spin-liquid ground state.

In this limit, one can generically show that the contribution of the perturbation $\tilde{g} \,\delta u \int\dd[d]{x} \mathcal{O}_\Delta(x)$ gives an energy gain
\begin{equation}
    \delta \varepsilon = - \frac{1}{2}(\tilde{g}\, \delta u)^2 \int\dd[d]{x} \langle \mathcal{O}(x)\mathcal{O}(0)\rangle.
\end{equation}
The result of this integral depends on the space-time dimension $d=D+1$ and the boundary conditions imposed, as well as the precise form of the correlation function.
In \cite{seifert2024}, it was shown that in the case of the U(1) Dirac spin liquid, evaluating this integral at finite temperature gives $\delta \varepsilon = - c_\Phi^\beta\, \beta^{3-2\Delta_\Phi}(\tilde{g}\,\delta u)^2$ which implies $(T_{\mathrm{SP}}/J_1)^{3-2\Delta_\Phi} = c_\Phi^\beta \, \tilde{g}^2/K$. Because we do not know the mapping between the QED$_3$ field theory and lattice parameters ($\tilde{g}$ and $g$), this field theory approach is not able to predict the numerical value of critical temperatures and we must therefore turn to numerical simulations to measure it. In this case, we use zero-temperature simulations on a finite-$L$ torus geometry which acts as an effective temperature $\T = 1/L$ and find $(\T_{\mathrm{SP}}/J_1)^{3-2\Delta_\Phi} = c_\Phi^L \,\tilde{g}^2/K$. Although $c_\Phi^L$ and $c_\Phi^\beta$ are not equal, we assume they are related by an order-one constant to justify considering this as an effective temperature.

In the 1D case \cite{orignac2004} performed this same finite-temperature calculation to extract a numerical value of the transition temperature $T_{\mathrm{SP}} / J_1 = 1.01 \lambda$. 
This calculation neglects the effects of logarithmic corrections, which would act to suppress $T_{\mathrm{SP}}$.
Note that in 1D, one may repeat this calculation on a zero-temperature system with periodic spatial boundary conditions. The resulting $\mathbb{R}\times S^1$ geometry of the space-time manifold is identical to the finite-temperature $L=\infty$ case of \cite{orignac2004}. One may then use the same conformal mappings to evaluate correlation functions and consequently finds that the effective critical temperature is equal to the true critical temperature of \cite{orignac2004}; that is, $\T = T$ in 1D.

\subsubsection{Scaling Ansatz}
In this work we consider the response of CFTs on finite-system sizes to relevant perturbations.
In order to describe the results beyond the weak-coupling limit and assess compatibilitu with the CFT, we leverage a scaling Ansatz $f(\tilde{g} \, \delta u, 1/L)$. This assumes that the ground state energy of the spin system $f$ depends on the energy scale introduced by distortion $\delta u$ and system size $L$. This energy density (energy per spacial area) has scaling dimension dimension $[f]=d$; furthermore we know the dimensions $[L]=-1$, and the coupling $[\tilde{g} \, \delta u] = d-\Delta$ by power counting.
Therefore, we can constrain how the function $f$ rescales under $x\to bx$, and find
\begin{equation}
    f(\tilde{g} \, \delta u, 1/L) \to 
    f(b^{-(3-\Delta_\Phi)} \,\tilde{g} \, \delta u, 1/bL)
    = b^{-3}f(\tilde{g} \, \delta u, 1/L).
\end{equation}
Choosing $b=L^{-1}$ we recover
\begin{equation}
    L^{3} f(\tilde{g} \, \delta u, 1/L) = f(\tilde{g} \, \delta u\, L^{3-\Delta_\Phi}, 1)
    \equiv \tilde{\mathcal{F}}(\tilde{g} \, \delta u\, L^{3-\Delta_\Phi}),
\end{equation}
where $\tilde{\mathcal{F}}(x)$ is a universal scaling function. This shows that plotting $L^{d} f$ against $L^{d-\Delta} $ should produce a data collapse.
The function behaves as $\tilde{\mathcal{F}}(x) = B x^{d/(d-\Delta)}$ when $x\gg1$ and conformal symmetry is restored. We refer to this as the strong-coupling regime.
In the opposite limit (of small $g$), it goes as $\tilde{\mathcal{F}}(x) = A x^2$ and produces a response compatible with the weak-coupling limit described in the previous section. 
This Ansatz is applied to produce the data collapses shown in Fig.~\ref{fig:collapses} of the main text.

\subsection{Dynamical phonons}

\subsubsection{Wave functions for spins and phonons}

To pursue a full quantum analysis of the spin-phonon problem, we turn to study the Hamiltonian of Eq.~\eqref{eq:spinphonon}, which, in addition to accounting for the potential energy cost of lattice distortions, also includes the phonons kinetic energy, providing them with quantum dynamics. Specifically, the free phonon part of the Hamiltonian~\eqref{eq:spinphonon} describes a set of uncoupled harmonic oscillators, i.e. Einstein phonons. A full quantum treatment of the problem is achieved by the definition of a variational wave function for spins and phonons
\begin{equation}    |\Psi_{\mathrm{sp}}\rangle=\mathcal{J}_{\mathrm{sp}} |\Psi_{\mathrm{p}}\rangle \otimes |\Psi_{\mathrm{s}}\rangle \,.
\end{equation}
This ansatz is the product of a spin state, $|\Psi_{\mathrm{s}}\rangle$,  a phonon state, $|\Psi_{\mathrm{p}}\rangle$, and a spin-phonon Jastrow factor $\mathcal{J}_{\mathrm{sp}}$~\cite{ferrari2020a,ferrari2021,ferrari2024}. The spin wave function is a Gutzwiller-projected fermionic state supplemented by a (long-range) spin-spin Jastrow factor, namely ${|\Psi_s\rangle=J_{\rm s}\mathcal{P}_G |\Phi_0\rangle}$, with
$J_s=\exp(\sum_{i,j} v_{i,j} S_i^z S_j^z)$. The variational parameters of $J_s$, $v_{i,j}=v_{j,i}$, depend only on the distance between sites in the undistorted lattice, i.e. $\|\bm{r}_i-\bm{r}_j\|$. As previously discussed, the fermionic state $|\Phi_0\rangle$ is taken to be the ground state of a certain fermionic Hamiltonian $\mathcal{H}_0$ [see Eq.~\eqref{eq:aux_H0}]. The phonon part of the variational Ansatz is a coherent state. Its amplitude on a generic element of the phonon Hilbert space (labelled by the local displacements) reads
\begin{equation}
 \langle \{\bm{u}_j\} | \Psi_{\mathrm{p}}\rangle =\prod_i \exp(-\frac{m \omega}{2} \|\bm{u}_i-\bm{z}_i\|^2)\,,
\end{equation}
i.e. it is a product of Gaussians whose centers are controlled by the variational parameters $\bm{z}_i\in \mathbb{R}^D$. A nonzero $\bm{z}_i$ upon optimization is associated with a finite displacement of site $i$. Finally, the variational state $|\Psi_{\mathrm{sp}}\rangle$ is completed by the Jastrow factor
\begin{equation}
\mathcal{J}_{\mathrm{sp}}= \exp\Big[ \sum\nolimits_{i,j}    \bm{w}_{i,j}  \cdot(\bm{u}_i-\bm{u}_j)  S^z_i S^z_j \Big]\,,
\end{equation}
which connects spin and phonon degrees of freedoms. The variational parameters of $\mathcal{J}_{\mathrm{sp}}$ depend only on the distance $\|\bm{r}_i-\bm{r}_j\|$, but are antisymmetric under the swap of lattice sites, namely $\bm{w}_{i,j}=-\bm{w}_{j,i}$. Note that $\bm{w}_{i,j}\in \mathbb{R}^2$ ($\in \mathbb{R}$) for the 2D (1D) spin-phonon problem. Details on the implementation of the Monte Carlo procedure in the Hilbert space of spins and phonons are given in~\cite{ferrari2024}.

\subsubsection{Dynamical phonons: Results}

We use the variational Ansatz $|\Psi_{\mathrm{sp}}\rangle$ to assess the ground state properties of the spin-phonon Hamiltonian~\eqref{eq:spinphonon}. Let us first discuss the case of the $J_1$--$J_2$ model on the triangular lattice coupled to quantum phonons. As already mentioned, at $J_2/J_1=1/8$ and in absence of spin-phonon interactions, i.e. $g=0$, the optimal wave function is the DSL state. When the spin-phonon interaction is turned on, we allow the variational state to (potentially) gain energy by creating lattice distortions. Concretely, we perform an unbiased search of the possible instabilities towards lattice distortions, following the same strategy adopted in \cite{ferrari2024}, i.e. considering different supercells which can accommodate different patterns of sites displacements. These supercells determine the periodicity of the wave functions $|\Psi_p\rangle$ and $|\Psi_s\rangle$: for $|\Psi_p\rangle$, this means that we independently optimize the $\bm{z}_i$ parameters of each site within the supercell; analogously, for $|\Psi_s\rangle$, we start from the ${\cal H}_0$ auxiliary Hamiltonian of the U(1) DSL state and we independently optimize all the hopping parameters inside the supercell. We experiment with supercells of different shape and size, up to a maximum of $24$ sites per supercell. This approach enables an unbiased exploration of various potential Peierls orders.

The first important observation of our analysis is that for small values of the spin-phonon coupling $g$ the system remains in the DSL phase and no distortions are observed. Thus, the spin liquid state proves to be stable to the Peierls mechanism when the quantum nature of lattice distortions is properly accounted for. On the other hand, for spin-phonon interactions larger than a certain critical value $g_c$, we observe a phase transition towards the $12$-sites VBS order associated with the $\bm{X}_\alpha$ phonon mode. It is worth emphasizing that here, contrary to the previous analysis of the response to static distortions, the $12$-sites VBS pattern is not imposed \textit{a priori}, but emerges spontaneously from the unbiased optimization of the variational parameters and the inspection of different supercells. Furthermore, the distortion pattern is compatible with Eq.~\eqref{eq:dist_2d} with $\delta u>0$ [Fig.~\ref{fig:supp_class}(b)], consistent with the expected behavior based on the static results of Fig.~\ref{fig:supp_class}(c). Similarly to what has been previously discussed in the static limit, also here the signs of the hopping parameters defining the optimal $|\Psi_s\rangle$ state maintain the structure of the DSL Ansatz, although their absolute values break the translational symmetry of the triangular lattice. We estimate the DSL-VBS critical point $g_c$, as a function of the phonon frequency $\omega$, by comparing the variational energy of the VBS state to the one of a DSL spin-phonon state in which only the Jastrow factors are optimized, while the parameters $\bm{z}_i$ are set to zero for all sites and the $\mathcal{H}_0$ Hamiltonian is the one of the pristine DSL Ansatz. Collecting the data for different system sizes $L$, we can then perform the data collapse analysis of Fig.~\ref{fig:g_scaling}, guided by the scaling Ansatz discussed in the following paragraph.

\subsubsection{Scaling Collapse}
The continuum action of the dynamical phonons [written on the lattice in Eq.~\eqref{eq:spinphonon}] is 
\begin{equation}
        S_{\mathrm{ph}}[\bm{u}] = \int \dd[d]{x} m \left[ \| \partial_\tau \bm{u} \|^2 + \omega^2 \|\bm{u}\|^2 \right]\, .
\end{equation}
As before, we identify $K = m \omega^2$ as the lattice stiffness.
Focusing on the QED$_3$-lattice model, the coupled theory [a dynamic form of \eqref{eq:qed3_lattice_coupling}] is then written as
\begin{equation}
    S_{\mathrm{QED}_3} + S_{\mathrm{ph}}[\bm{u}] + \tilde{g} \int\dd[3]{x} [\mathcal{V}(x)\, u(x) + \mathrm{h.c.}],
\end{equation}
where $u(x)$ is the same longitudinal component of the vector field $\bm{u}(x)$ as in the static case.
By integrating out the free phonons exactly, one is left with an effective retarded monopole-monopole interaction, controlled by the phonon propagator $G(\tau-\tau')$ as follows
\begin{equation}
    S_{\mathrm{I}} = \frac{\tilde{g}^2}{m} \int\dd[2]{\bm{x}}\!\int\dd{\tau}\dd{\tau'}
    \mathcal{V}(\bm{x},\tau)\, G(\tau-\tau')\mathcal{V}^\dagger(\bm{x},\tau') \, .
\end{equation}
This operator drives the U(1) DSL--VBS phase transition, with the ordered phase occurring upon sufficiently large interaction strength.
The field-theoretic coupling strength is $\tilde{g}^2/m$ and has scaling dimension $5-2\Delta_\Phi$ by power counting.
The other important parameter in the system is $\omega$ (scaling dimension 1), which sets the interaction length scale in imaginary time and also controls the propensity of the monopoles to precipitate VBS order. In the antiadiabatic limit $\omega\to\infty$ there is no retardation and the DSL is stable; in the adiabatic limit, an infinitesimal coupling causes VBS ordering.

To make progress in interpreting our numerical results, we make a scaling Ansatz for the \emph{critical} coupling strength $\tilde{g}_c^2/m$ at which the transition occurs. This is a function only of the phonon frequency and the finite size, $\eta_c=[\tilde{g}_c^2/m](\omega, 1/L)$, which we constrain under rescaling $x\to bx$, finding
\begin{multline}
    [\tilde{g}_c^2/m](\omega, 1/L) \to 
   [\tilde{g}_c^2/m](b^{-1} \omega, 1/bL)\\
    = b^{-(5-2\Delta_\Phi)}[\tilde{g}_c^2/m](\omega, 1/L).
\end{multline}
Choosing $b=L^{-1}$ we recover
\begin{equation}
    L^{5-2\Delta_\Phi}[\tilde{g}_c^2/m](\omega, 1/L) = [\tilde{g}_c^2/m](\omega L, 1)
    \equiv \tilde{\mathcal{G}}(\omega L),
\end{equation}
where $\tilde{\mathcal{G}}(x)$ is a universal scaling function.
One can show in the large-$\omega$ regime that the critical coupling is finite; here we have $\tilde{\mathcal{G}}(x) = a_2 x^{5-2\Delta_\Phi}$ and so
$\lambda_c=\tilde{g}_c^2/(m\omega^2)=\tilde{g}^2/K = a_2 \omega_c^{3-2\Delta_\Phi}$.
In the small-$\omega$ limit, the scaling form goes as $\tilde{\mathcal{G}}(x) = a_1 x^2$ and so $\lambda_c = a_1 \T_c(\lambda)^{3-2\Delta_\Phi}$ defines the critical temperature for the U(1) DSL as a function of $\lambda = \tilde{g}^2/K$, where we take $\omega\to 0$ with $K$ fixed.
This Ansatz is applied in 2D to produce the data collapse shown in Fig.~\ref{fig:g_scaling}(b) of the main text.
While no closed analytical expression for the full scaling form is available, we find that the whole range of the collapsed data is well described by the heuristic fit
\begin{equation}
    \mathcal{H}(x)_{\mathrm{fit}} = a_1 x^2 + a_2 x^{d+3-2\Delta_\Phi}
\end{equation}
with parameters $a_1 = 0.58(2)$, $a_2 = 0.138(5)$ and $\Delta_\Phi = 1.23(3)$. This reproduces the expected asymptotic regimes and interpolates between them, generating the smooth curve plotted in the phase diagram of Fig.~\ref{fig:phasediag}. Here, the continuous line implicitly defined through
\begin{equation}
    \lambda = \tilde{g}_c^2/(m\omega^2) = \mathcal{H}(\omega/\T)_{\mathrm{fit}} / \omega^2 = a_1 \T_c^{\ 3-2\Delta_\Phi} + a_2 \omega^{3-2\Delta_\Phi}\, ,
\end{equation}
for various $\lambda$.

\end{document}